\documentclass[aps,prc,superscriptaddress]{revtex4-2}
\usepackage{CJKutf8}
\usepackage{amssymb}
\usepackage{amsmath}
\usepackage{diagbox}
\usepackage{makecell}
\usepackage{tikz}
\usepackage{color}
\usepackage{graphicx} 
\usepackage{multirow} 
\usepackage{makecell}
\usepackage{ulem}
\usepackage[hidelinks]{hyperref}
\def\be{\begin{equation}}
\def\ee{\end{equation}}
\def\bea{\begin{eqnarray}}
\def\eea{\end{eqnarray}}

\usepackage[table]{xcolor}
\newcommand{\bit}[1]{\textbf {\textit #1}}

\begin{document} 

\title{A Halo: The Trigger to a New Era of Nuclear Correlations}

\author{Hiroyuki Sagawa}  
\email[]{sagawa@ribf.riken.jp}
\affiliation{Institute of Theoretical Physics, Chinese Academy of Sciences, Beijing 100190, China}
\affiliation{RIKEN Nishina Center for Accelerator-Based Science, Wako 351-0198, Japan} 
\affiliation{Center for Mathematics and Physics University of Aizu, Aizu Wakamatsu, Fukushima 965-0001, Japan}
\author{Xiao Lu}  
\email[]{luxiao@itp.ac.cn}
\affiliation{Institute of Theoretical Physics, Chinese Academy of Sciences, Beijing 100190, China}
\author{Shan-Gui Zhou}  
\affiliation{Institute of Theoretical Physics, Chinese Academy of Sciences, Beijing 100190, China} 
\affiliation{School of Physical Sciences, University of Chinese Academy of Sciences, Beijing 100049, China} 
\affiliation{School of Nuclear Science and Technology, University of Chinese Academy of Sciences, Beijing 100049, China}  

\date{\today}

\begin{abstract}
In this contribution to the Halo-40 Proceedings, we discuss two topics regarding halo phenomena: The first is the pairing anti-halo effect on the neutron radius of halo nuclei and its restoration due to the coupling to the continuum; the second is the soft dipole excitation of deformed halo nuclei. 
We demonstrate the importance of Hartree-Fock-Bogoliubov and the relativistic Hartree–Bogoliubov theory in continuum for properly taking into account the halo nature of extended wave functions in calculations of neutron radii, as well as the soft  dipole excitations of halo nuclei.  It was shown that the anti-halo effect is very sensitive to the continuum coupling induced by Bogoliubov-type quasi-particles, which largely cancels the anti-halo effect on the neutron radius.  The soft dipole excitations of deformed halo nuclei $^{31}$Ne and $^{37}$Mg are discussed within the deformed Woods-Saxon model.  We point out that the sharp peak just above the threshold in the dipole response is created by the halo effect, and its strength can be used to identify the magnitude of deformation and the halo configuration in the Nilsson level scheme.

\end{abstract}


\maketitle
\section{Introduction}

Recent progress in radioactive beams has opened a new era in nuclear structure and reactions, extending the field into regions close to and beyond the drip lines of the nuclear chart.  
In fact, the physics of neutron-rich and proton-rich nuclei has become one of the main subjects of nuclear physics research~\cite{TANIHATA2013215,HTS13,Zhou2017_PoS}.
In the 2020s, a new generation of RI beam facilities has come online or will soon be in operation worldwide, such as RIBF at RIKEN (Japan)~\cite{RIBF2007}, FRIB at MSU (USA)~\cite{FRIB2019}, FAIR at GSI (Germany)~\cite{FAIR2019}, SPIRAL2 at GANIL (France)~\cite{SPIRAL22022}, SPES at LNL (Italy)~\cite{SPES2020}, HIAF at IMP (China)~\cite{Zhou2022HIAF}, and RAON at IBS (Korea)~\cite{RAON2025}. 

Until the mid-1980s, nuclear physics research was primarily focused on the 251 known stable nuclides. In contrast, the number of unstable nuclei observed in nature is about 100, and more than 3,000 nuclei have been created in laboratories~\cite{Kondev2021NUBASE}. Thus, unstable nuclei may offer more opportunities to explore new structures and reactions in laboratory experiments.

\begin{figure}[tb]
\vspace{-0.5cm}
\begin{center}
\includegraphics[width=10 cm]{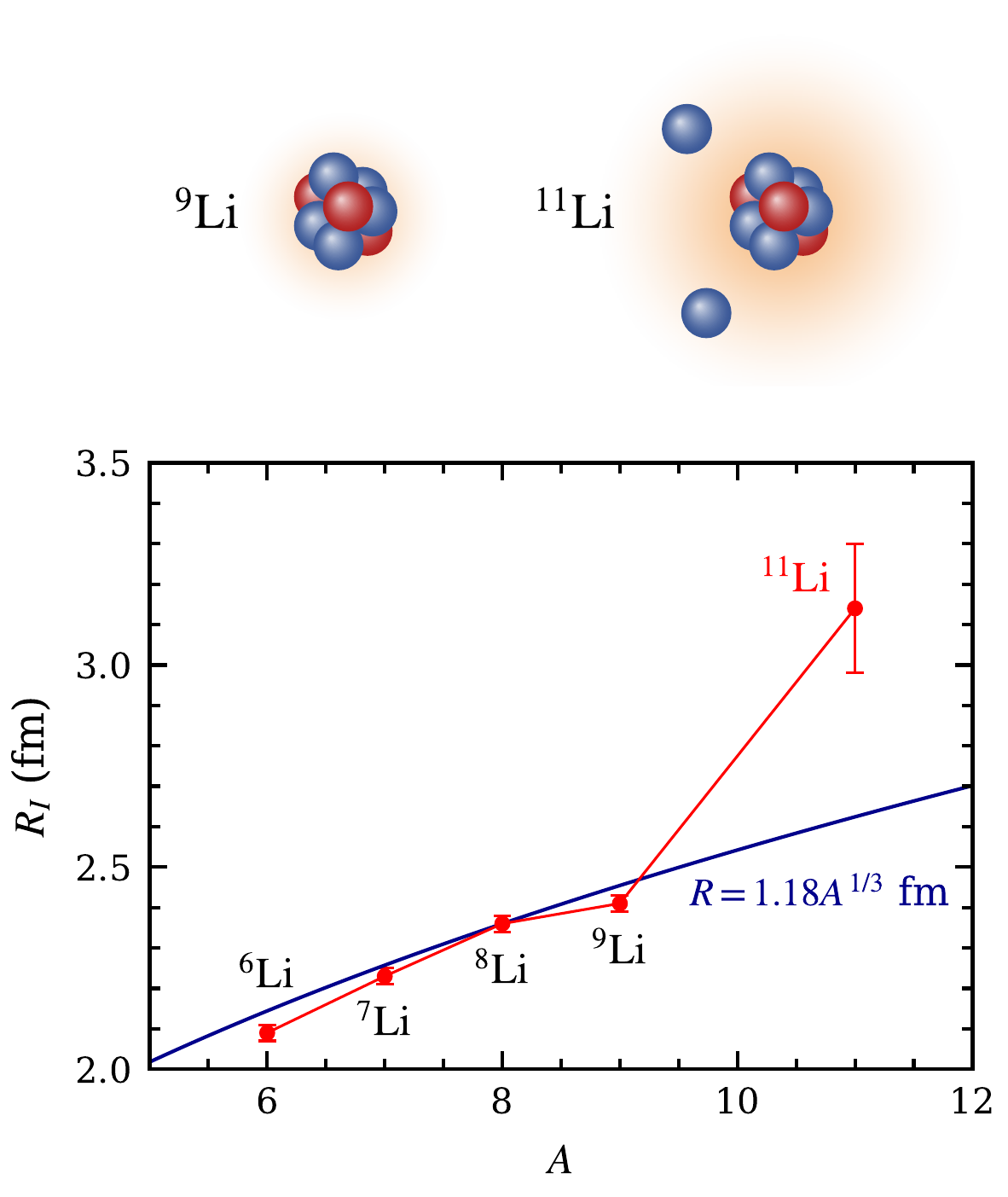} 
\caption{The matter radii of Li isotopes, deduced from the measured interaction cross sections with a carbon target at 790 MeV/nucleon~\cite{Tani85}.  The solid line represents the systematic trend for stable nuclei, given by $R=1.18A^{1/3}$fm, where $A$ is the mass number.  }
\label{fig:1}       
\end{center}
\end{figure}

The breakthrough in the physics of unstable nuclei was the experimental discovery of an anomalously large matter radius in the $^{11}$Li nucleus by Tanihata {\it et al.}~\cite{Tani85} through measurement of interaction cross sections.  
On average, nuclear radii increase with the mass number $A$ of a nucleus as $A^{1/3}$, reflecting the saturation property of the nuclear interaction as seen in Fig. \ref{fig:1}.  It was therefore a big surprise that the matter radius of $^{11}$Li deviates significantly from the $A^{1/3}$ law, while other Li isotopes follow it quite precisely. This discovery has raised new concepts in nuclear physics, such as ``neutron halo'', ``di-neutron'' correlations,  and Borromean nuclei, of which any two-body subsystem is not bound (two neutrons or $^{10}$Li), though the three-body system as a whole is bound ($^{11}$Li). 
A straightforward interpretation of this large matter radius comes from its small two neutron separation energy of $^{11}$Li, $S_{2n}=378 \pm 5$ keV~\cite{Bachelet08}.  
This structure has been referred to as a ``halo'', in which the density distribution of valence neutron(s) extends far beyond the core nucleus. 
This halo phenomenon has also been observed in several other loosely bound nuclei, such as $^{11}$Be~\cite{1988Beisotopes}, $^{19}$C~\cite{Ozawa2001C19},  $^{6}$He~\cite{1985Heisotopes},  $^{8}$He~\cite{1985Heisotopes}, and $^{14}$Be~\cite{1988Beisotopes}, through measurement of interaction cross section. 

Immediately following the observation of the large matter radius of the halo nucleus $^{11}$Li, the extended structure of the valence neutron wave functions was further confirmed by the observation of the narrow momentum distributions of the $^{9}$Li core resulting from the breakup of $^{11}$Li~\cite{Kobayashi88}. The halo structure also gives rise to a large electric dipole ($E1$) transition strength just above the neutron threshold in the Coulomb dissociation cross sections, which is called ``soft dipole'' excitation.

Since the first observation of a halo nucleus in 1985, many halo nuclei have been experimentally identified; for examples, the $1p$ halo nuclei $^8$B~\cite{1996B8} and $^{12}$N~\cite{1998N12}, the $1n$ halo nuclei {$^{11}$Be~\cite{Nakamura199411Be}, $^{15}$C~\cite{Fang2004C15}, $^{19}$C~\cite{Bazin1995C19},} $^{31}$Ne~\cite{Nakamura2009,Nakamura2014}, and $^{37}$Mg~\cite{mg37,mg371}, as well as the $2p$ halo nucleus $^{17}$Ne~\cite{2003Ne17}, and the $2n$ halo nuclei {$^{6}$He~\cite{1985Heisotopes}, $^{14}$Be~\cite{Labiche200114Be},} $^{17}$B \cite{17B02,17B21}, $^{19}$B \cite{19B20}, $^{22}$C \cite{Tanaka2010}, and $^{29}$F \cite{29F20}.  
An interesting question in halo evolution is how shell structure, pairing correlations, and deformation effects contribute to their formation,  and in which regions of the nuclear chart they occur, especially in relation to nuclear structure near the drip lines and astrophysical reaction processes.

The breakthroughs in new structures and reactions from the observation of halo nuclei gave rise to the following topics related to exotic phenomena~\cite{HTS13,Zhou2017_PoS}:
\begin{itemize}
\setlength{\itemindent}{1cm}
\item  halo and giant halo/skin structures
\item  shell evolution near the drip lines
\item  di-neutron and di-proton correlations
\item  {Bose-Einstein condensation (BEC)-BCS crossover} 
\item  deformed halo and shape decoupling
\item soft dipole excitation
\item anti-halo and its restoration by continuum coupling
\item new radioactivities: $1p$, $2p$, and $2n$ emissions
\item clustering effects in loosely bound systems.
\end{itemize}
These exotic phenomena can be accessed experimentally by various reaction processes~\cite{TANIHATA2013215}:
\begin{itemize}
\setlength{\itemindent}{1cm}
\item Coulomb breakup reactions
\item nuclear breakup reactions
\item nuclear transfer and knockout reactions
\item  direct  two-nucleon decays
\item n-n correlation measurements
\item charge exchange reactions.
\end{itemize}

In this paper, we review briefly two selected topics on halo phenomena: One is the anti-halo effect and its restoration due to strong continuum coupling, and another is the deformed halo and soft dipole excitation. The anti-halo effect originates from the asymptotic behavior of Hartree-Fock-Bogoliubov (HFB) or Relativistic Hartree-Bogoliubov (RHB) wave functions. Namely, if the gap parameter $\Delta_i^{(\rm can)}$ remains finite as the single-particle energy $\varepsilon_i^{(\rm can)}$ approaches zero, pairing correlations will substantially suppress the extreme spatial extension of the halo wave function, which would otherwise occur in a Hartree-Fock (HF) field, thereby preventing the divergence of the root-mean-square (rms) radius. This is referred to as the anti-halo effect due to pairing correlations~\cite{BDP00}.  
This effect can be observed experimentally~\cite{Takechi12,mg371} by systematically studying the reaction cross sections of given isotopic chains, {\it e.g.,} Ne or Mg isotopes~\cite{HS11,HS12,Sasabe13,MY14}. On the other hand, the pairing effect will induce strong coupling to the continuum, especially if nucleons are loosely bound. That is, the pairing correlation scatters a nucleon pair to continuum states, thus increasing the rms radius and enhancing the dipole response. These two effects compete with each other, and the latter may become important in certain nuclei~\cite{Chen2014,mg371}.  We discuss this problem in terms of the soft dipole response in a deformed halo nucleus $^{37}$Mg, using the deformed  relativistic Hartree-Bogoliubov theory in continuum (DRHBc) method. 

Theoretically, deformed halo nuclei have been studied within various frameworks, including the deformed Woods-Saxon model  \cite{Hamamoto2005,Hamamoto2012,H10,Xiao2025}, the particle-rotor model \cite{UHS11,Urata2012,Urata2013}, the shell model \cite{Otsuka93,Kuo97}, the cluster model \cite{DESCOUVEMONT1999}, the DRHBc theory~\cite{Zhou10,li12,SUN18,zhang19,SUN20,SUN21,SUN20212072,Zhong2022,ZhangKY23,ZHANG2023138112,plbzhang2024}, the non-relativistic HFB approach~\cite{pei131,pei132,Nakada18,Kasyua2021,peisy21}, and the anti-symmetrized molecular dynamics model~\cite{Takatsu2023}. We adopted a deformed Woods-Saxon model to study the intrinsic deformation effects on the dipole response of $^{31}$Ne and $^{37}$Mg in Ref. \cite{Xiao2025}. Recently we have extended our study of deformed halos by employing the DRHBc theory.  In this contribution to the Halo-40 Proceedings, we will present some new results on soft dipole excitations obtained by using the DRHBc theory with the PC-F1 effective density functional~\cite{Burvenich2002Phys.Rev.CNuclear}. 
 
This paper is organized as follows. Section \ref{sec:2a} is devoted to the theoretical models adopted in the study.  
The anti-halo effect and its restoration due to the coupling to the continuum are discussed in Section \ref{sec3}. In Section \ref{sec4}, we discuss the dipole response within a deformed Woods-Saxon model, as well as the anti-halo effect and its restoration as reflected in the dipole response calculated using the DRHBc theory. Finally, a summary and future perspectives are provided in Section \ref{sec5}. 

\section{The DRHBc Theory}
\label{sec:2a}

For exotic nuclei, the conventional BCS approach proves to be a poor approximation, as it fails to capture the features of loosely bound wave functions~\cite{B00,DFT84,DNW96,Meng2006}. Moreover, coupling to the continuum is a critical issue when studying nuclei near the drip lines. {Therefore, a self-consistent framework that treats pairing correlations and continuum effects on an equal footing is required.}
In this section, we briefly summarize the DRHBc theory; the details can be found in Refs. \cite{Zhou10,li12}. 
 
In DRHBc model, the general Bogoliubov transformation for the quisi-particle $\alpha_k^\dagger$ is given by:
\begin{equation}\label{eq:Bogoliubov}
\alpha_k^\dagger = \sum_l U_{lk} a^\dagger_l + V_{lk} a_l,    
\end{equation}
where the matrices $U$ and $V$ generalize the coefficients $u$ and $v$ in the BCS model, and $a_l$ and $a^\dagger_l$ are the annihilation and creation operators of bare particle, respectively.  
Including the annihilation operators $\alpha$, this becomes:
\be
\left( \begin{array}{c} \alpha \\ \alpha^\dagger \end{array} \right) = {\cal W}^\dagger
 \left(\begin{array}{c} a \\ a^\dagger \end{array} \right),
 \,\,\, {\rm with} \,\,\  \cal W=
 \left( \begin{array}{cc} U & V^*\\ 
V  & U^* \end{array} \right).
\ee
One would like the canonical anticommutation relations to be preserved, also in the case of the general Bogoliubov transformation. 
Since the matrix $\cal W$ is unitary so that ${\cal W}^\dagger = {\cal W}^{-1}$.   The inverse transformation can be expressed as:
\be 
\left( \begin{array}{c} a \\ a^\dagger \end{array} \right) = {\cal W}
 \left(\begin{array}{c} \alpha  \\ \alpha ^\dagger \end{array} \right),
\ee
The quasi-particle state can be expressed in the coordinate space representation as:
\be  \label{eq:HFB}
\phi_k(x)=\left(\begin{array}{c} U_k(x) \\ V_k(x)  \end{array} \right) =\left(\begin{array}{c} U_{ik}\psi_i(x) \\ V_{ik}\psi_i(x)\end{array} \right), 
\ee
where the quasiparticle wave functions $U_k$ and $V_k$ in Eq. \eqref{eq:HFB} 
are Dirac spinors. Each of them is expanded in terms
of spherical Dirac spinors $\psi_{n\kappa m}({\bit r}sp)$ with the eigenvalues $\varepsilon_{n\kappa}$, {\it i.e.}, 
 the solution of a Dirac equation $h^{(0)}_D$ containing spherical scalar and vector potentials.   
If we neglect the Fock terms as it
is usually done in the covariant density functional theory, the
Dirac Hartree Bogoliubov (RHB) equation for the nucleons
reads:\\
\begin{eqnarray}
\label{eq:HFBd}
 \int d{\bit r}^\prime 
\left( \begin{array}{cc} h_D({\bit r},  {\bit r}^\prime, ) -\lambda& \Delta({\bit r},  {\bit r}^\prime ) \\ -\Delta^*({\bit r}, {\bit r}^\prime) & -h_D^*({\bit r},  {\bit r}^\prime) +\lambda \end{array} \right) 
\left( \begin{array}{c} U_k({\bit r}^\prime ) \\ V_k({\bit r}^\prime ) \end{array} \right)  
= E_k
\left( \begin{array}{c} U_k({\bit r}) \\ V_k({\bit r}) \end{array} \right),
\end{eqnarray}
where $E_k$ is the quasi-particle energy and $\lambda$ is the chemical potential, and $h_D$ is the Dirac Hamiltonian. 
The pairing field is determined by the effective pairing interaction $V^{pp}$ and the pairing tensor $\kappa$,
\be
\Delta(\boldsymbol r,\boldsymbol r')
=
V^{pp}(\boldsymbol r,\boldsymbol r')
\kappa(\boldsymbol r,\boldsymbol r').
\ee

\section{Anti-halo effect and its restoration}\label{sec3}

\subsection{HFB model in the coordinate space}
The HFB method in the coordinate space representation is a useful approach to take into account the large spatial extension of wave functions with small binding energy~\cite{DFT84,DNW96,B00}. 
In the coordinate space representation,  the quasi-particle creation and annihilation operators are expressed through a Bogoliubov transformation:
\bea \label{B-trans}
\alpha_i^{\dagger}&=&\int dx\{u_i(x)\psi^{\dagger}(x)+v_i^*(x)\psi(x)\}, \nonumber \\
\alpha_i&=&\int dx\{u_i^*(x)\psi(x)+v_i(x)\psi^{\dagger}(x)\},   \quad \quad x=\{\boldsymbol{r}, \sigma \}
\eea
where $\psi^{\dagger}(x) (\psi(x))$ is the bare particle creation (annihilation)
operator at $x$.  The quasi-particle wavefunctions $u_i(x)$ and $v_i(x)$ are determined by 
the HFB equations~\cite{DFT84,DNW96,B00},
\bea
\left(\begin{array}{cc}
\hat{h}-\lambda & \Delta(\boldsymbol{r}) \\
\Delta(\boldsymbol{r}) & -\hat{h}+\lambda
\end{array}\right)\binom{u_i(\boldsymbol{r})}{v_i(\boldsymbol{r})}=E_i\binom{u_i(\boldsymbol{r})}{v_i(\boldsymbol{r})},
\label{HFB-eq}
\eea
where 
\begin{equation}
\hat{h}=-\frac{\hbar^2}{2m}\nabla^2 + V(\boldsymbol{r}), 
\end{equation}
is the mean-field Hamiltonian, $m$ being the nucleon mass. 
$V(\boldsymbol{r})$ and $\Delta(\boldsymbol{r})$ are the mean-field and pairing potentials, respectively, and $E_i$ is the quasi-particle energy. 

When the quasi-particle energy is smaller than the absolute value of the Fermi energy ($\lambda <0)$,  {\it i.e.}, $E_i<|\lambda|$, both the pairing wave functions $u_i(\boldsymbol{r})$  and  $v_i(\boldsymbol{r})$ are localized in space.
On the other hand, while the lower component $v_i(\boldsymbol{r})$ is always localized, the upper component $u_i(\boldsymbol{r})$ becomes non-localized if the quasi-particle energy $E_i>|\lambda|$. 
The asymptotic behavior of the HFB wave function for $E_i>|\lambda|$ can be determined by Eq. (\ref{HFB-eq}) in the limit $ V(\boldsymbol{r}), \Delta(\boldsymbol{r})\rightarrow 0$ as $\boldsymbol{r}\rightarrow \infty$:
\bea
u_i(\boldsymbol{r}) &\rightarrow& \frac{\sin(\beta_i r+\delta_i)}{r}, \label{HFB-u}\\
 v_i(\boldsymbol{r}) &\rightarrow & \frac{e^{-\gamma_i r}}{r},  \label{HFB-v}
\eea
where $\beta_i=\sqrt{2m(E_i+\lambda)/\hbar^2}$ and $\gamma_i=\sqrt{2m(E_i-\lambda)/\hbar^2}$. 
The normal density $\rho(r)$ and the abnormal density $\kappa(r)$ behave as:
\bea
\rho(r)&=&\sum_{E_i>0}| v_i(\boldsymbol{r}) |^2 \rightarrow 
\frac{e^{-2\gamma_{\rm min}r}}{r^2}, \\
\kappa(r) &=&\sum_{E_i>0} u_i(\boldsymbol{r}) v_i(\boldsymbol{r}) \rightarrow 
\frac{e^{-\gamma_{\rm min}r}}{r^2}, 
\eea
with $\gamma_{\rm min}=\sqrt{2m(E_{\rm min}-\lambda)/\hbar^2}$, where 
$E_{\rm min}$ is the lowest positive quasi-particle energy of the HFB equations. 

The pair potential $\Delta(\boldsymbol{r})$ generally exhibits a larger surface diffuseness than the mean field potential $V(\boldsymbol{r})$, and extends beyond it due to the non-localized nature of the upper component $u_i(\boldsymbol{r})$, which is coupled to the continuum spectra~\cite{DNW96}.  

\subsection{Anti-halo effect in HFB model}\label{sec3.1}
In the mean-field approximation without pairing correlations ({\it i.e.}, $\Delta(\boldsymbol{r})=0$), the halo structure originates from the occupation of a weakly-bound $l=0$ or $l=1$ orbit by valence nucleons near the threshold~~\cite{Sagawa93,Riisager}.  The asymptotic behavior of a single-particle wave function for the $s$-wave reads:
\begin{equation}
 \psi_i(r) \sim \frac{e^{-\alpha_i r}}{r},
\end{equation}
where $\alpha_i=\sqrt{2m|\varepsilon_i|/ \hbar ^2}$, and $\varepsilon_i$ is the HF or the Hartree single-particle energy. 
The mean square radius of this wave function is then evaluated as:  
\begin{equation}
 \langle r^2\rangle_{\rm HF}=\frac{\int r^2 |\psi_i(r)|^2 d\boldsymbol{r}}
{\int  |\psi_i(r)|^2 d\boldsymbol{r}} \propto 
   \frac{1}{\alpha_i^2}= \frac{\hbar^2}{2m|\varepsilon_i|},
\end{equation}
which diverges in the limit of vanishing separation energy, $|\varepsilon_i| \rightarrow 0$.  
It has been shown that this divergence occurs not only for $s$-wave but also 
for $p$-wave, although the dependence on $|\varepsilon_i|$ for $l=1$ is $\langle r^2\rangle_{\rm HF}\propto 1/\sqrt{|\varepsilon_i|}$~\cite{Riisager}.  
\begin{figure}[htb]
\begin{center} 
   \includegraphics[width=8 cm]{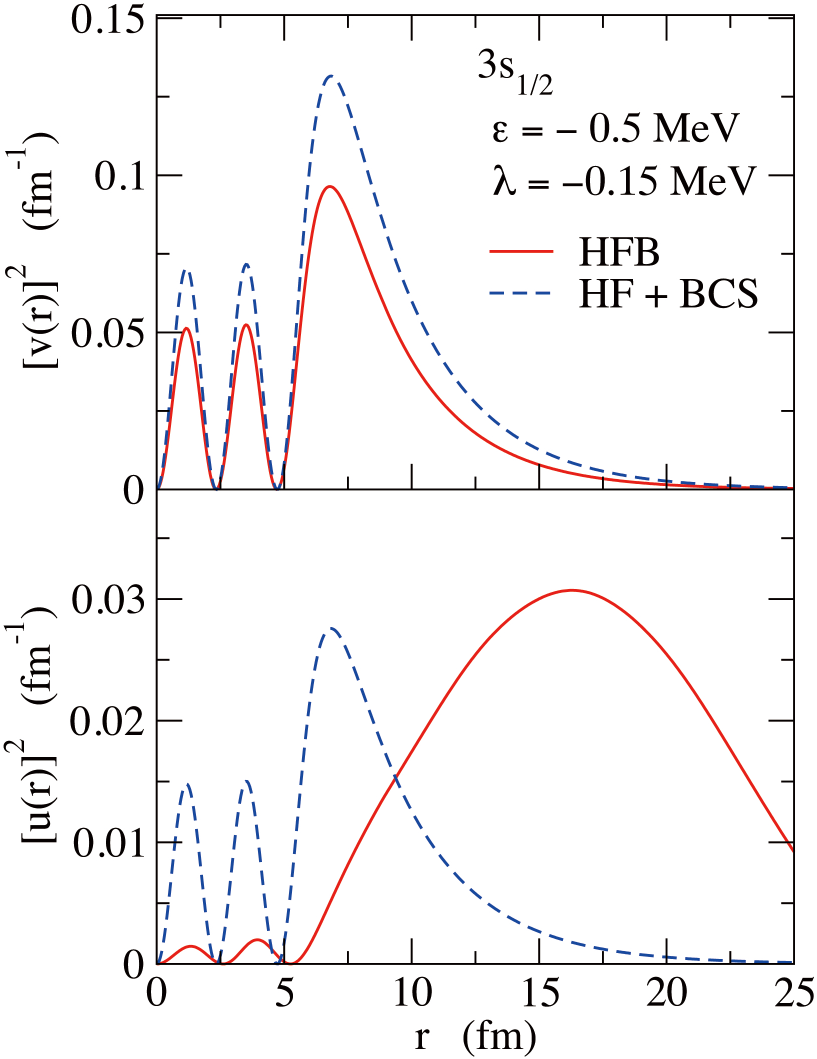} 
   \caption{The HFB wave functions with a binding energy in
     a HF potential of $\varepsilon=-0.5$ MeV and the Fermi energy of $\lambda=-0.15$ MeV.
     Since the quasi-particle energy 
     $E_i$ is larger than $|\lambda|$ for the loosely bound neutron,
     the  wave function $u_i$ is non-localized, while $v_i$ is localized.
     These wave functions are 
     compared with the BCS pair wave functions, which are obtained by
     multiplying factors to the HF wave functions.  
In the upper panel of the figure, the HFB and BCS wave functions $v_i$ show similar radial behavior, with the HFB wave function exhibiting some anti-halo effect. In the lower panel, the HFB wave function $u_i$ displays a typical non-localized behavior, in marked contrast to the BCS wave function, which remains always localized. Reproduced with permission from Ref. \cite{Sagawa2025}. Copyright (2025) by Springer Nature.
}
\label{fig:HFB}       
\end{center}
\end{figure}

In contrast, in the presence of the pairing correlations ({\it i.e., } $\Delta(\boldsymbol{r})\neq 0$), 
the lower component of the HFB (or Hartree-Bogolyubov, HB)  wave function, which is 
relevant to the density distribution, 
is given by Eq. (\ref{HFB-v}).
Here, the quasi-particle energy 
may be evaluated using the canonical basis $\psi_i^{\rm (can)}(\boldsymbol{r})$, which diagonalize the density matrix within HFB (HB) theory~\cite{DNW96}. Although the canonical basis is not an eigenstate of the HFB Hamiltonian, it provides a convenient means to analyze the structure of quasi-particle wave functions in terms of the BCS-like approximation. 
Notice that one can construct HFB (HB) quasi-particle states 
perturbatively starting from BCS states~\cite{pHFB}, and thus also from canonical basis states. 
In the canonical representation, the quasi-particle energies $E_i$ can be
approximately expressed as:
\begin{equation}
E_i=\sqrt{(\varepsilon_i^{\rm (can)} -\lambda)^2 +(\Delta_i^{\rm (can)})^2},
\label{EHFB} 
\end{equation}
where 
$\varepsilon_i^{\rm (can)}$ and 
$\Delta_i^{\rm (can)}$ {are the single-particle energy and the gap parameter in the canonical basis, respectively.}
In the zero binding limit, $\varepsilon_i^{(\rm can)} \sim 0$ and $\lambda  \sim 0$, the asymptotic behavior of $v_i(r)$ is determined by the gap parameter as:
\begin{equation}
  v_i(r)\propto  \frac{1}{r}\,
  \exp\left[\left(-\sqrt{\frac{2m}{\hbar^2}\Delta_i^{(\rm can)}}\right) r\right]. 
\label{HFB}
\end{equation}
The radius of the HFB wave function in the limit of small separation energy ($|\varepsilon_i^{(\rm can)}| \rightarrow 0$) is given by: 
\begin{equation}
 \langle r^2\rangle_{\rm HFB}
=\frac{\int r^2 |v_i(r)|^2 d{\boldsymbol{r}}}{\int  |v_i(r)|^2 d\boldsymbol{r}} \propto 
   \frac{1}{\beta_i^2} \rightarrow \frac{\hbar^2}{2m\Delta_i^{(\rm can)}}.  
\end{equation}
{If $\Delta_i^{(\rm can)}$ remains finite as $\varepsilon_i^{(\rm can)} \to 0$, the spatial extension of the halo wave function is substantially reduced by pairing correlations, preventing the rms radius from diverging. This is known as the pairing anti-halo effect~\cite{BDP00}.}   

An example of the HFB wave function for the $3s_{1/2}$ orbit with a HF energy of $\varepsilon=-0.5$ MeV is shown in Fig.~\ref{fig:HFB}. In this calculation, the Fermi energy is set to be $\lambda=-0.15$ MeV. One can clearly see that the tail of $v(r)$ for the HFB wave function is suppressed by the pairing effect compared to that of the HF+BCS model, where the radial dependence of the wave function is inevitably determined by the corresponding HF wave function.

\begin{figure}[ht]
\begin{center} 
\includegraphics[clip,width=8.cm,bb= 0 0 400 530]{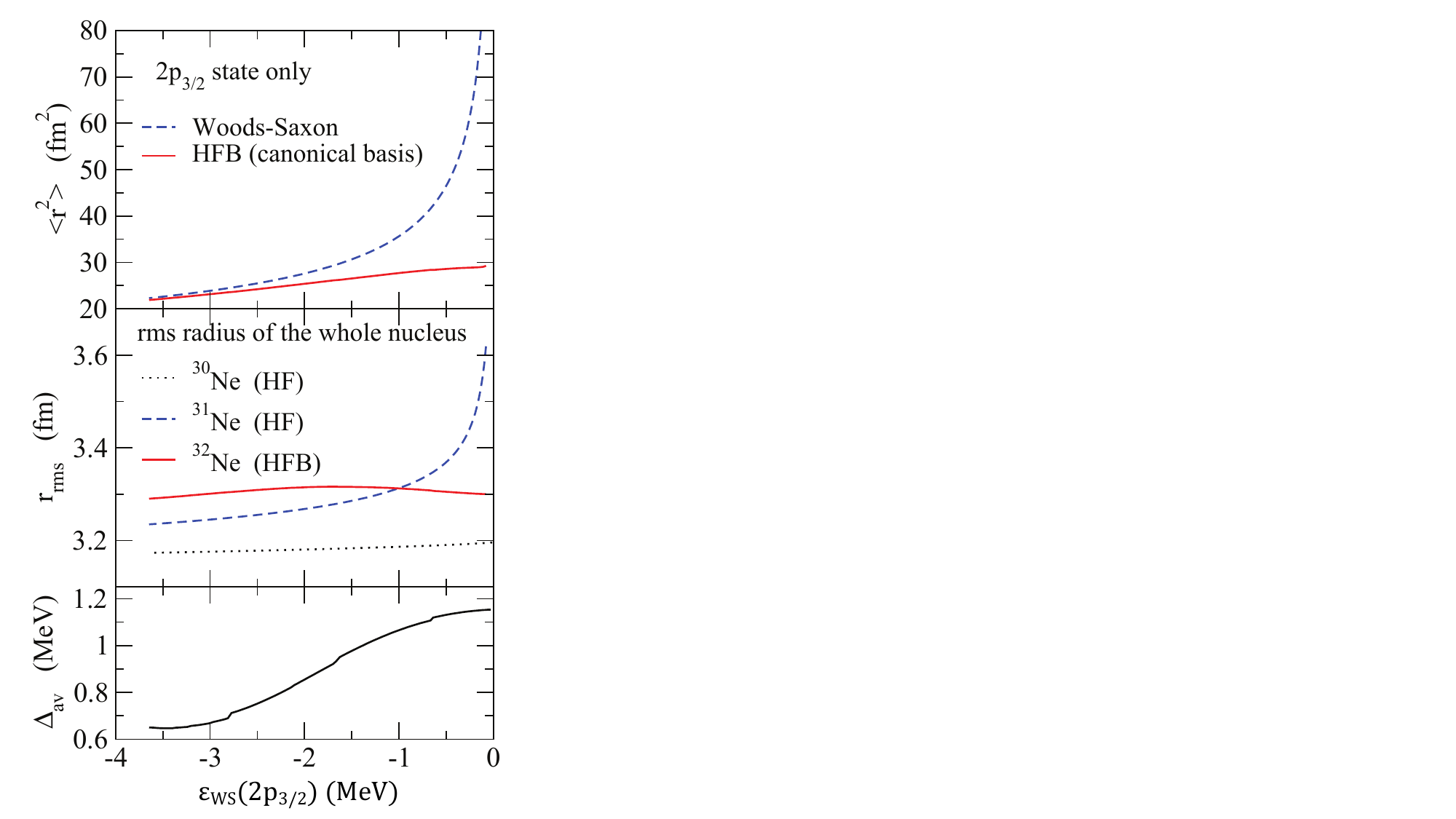}
\caption{
The mean square radii and the average pairing gap as a function of 
the single-particle energy $\varepsilon_{\rm WS}$ in a Woods-Saxon mean-field 
potential. 
The top panel shows the mean square radius of the 2p$_{3/2}$ wave function 
with and without the pairing correlation, denoted by 
HFB and Woods-Saxon, respectively.
The middle panel shows the rms radii 
for $^{30}$Ne (the dotted line), $^{31}$Ne (the dashed line), 
and $^{32}$Ne (the solid line), obtained with the Hartree-Fock ($^{30}$Ne and 
$^{31}$Ne ) and the HFB ($^{32}$Ne) calculations. 
The bottom panel shows the 
results of the HFB calculations for the 
average pairing gap of $^{32}$Ne.   
Reproduced with permission from \cite{HS11}. Copyright (2011) by the American Physical Society. 
}

\label{ah:fig2}
\end{center}
\end{figure}

\subsection{Anti-halo effect vs continuum coupling}
\label{sec:5}

The shrinkage of the HFB wave function tail shown in Fig.~\ref{fig:HFB} is a direct manifestation of the ``anti-halo effect''. Despite having the same single-particle energy ($\varepsilon=-0.5$ MeV) as the HF state, the HFB wave function exhibits a more compact distribution. This effect can be inferred experimentally \cite{Takechi12,mg371} by systematically studying the reaction cross sections of given isotopic chains, {\it e.g.,} Ne or Mg isotopes \cite{HS11,HS12,Sasabe13,MY14}.

To investigate this possibility, mean-field calculations are performed by using a Woods-Saxon (WS) potential without and with the Bogoliubov pairing correlations. 
As an example, the 2$p_{3/2}$ state in $^{31}$Ne is selected. Although $^{31}$Ne is most likely a deformed nucleus~\cite{H10,UHS11}, for simplicity and to isolate the pairing effect, a spherical Woods-Saxon mean-field potential is assumed.  {More realistic self-consistent calculations based on the DRHBc model will be presented in the following section.} 
Notice that a Woods-Saxon potential with a {relatively} large diffuseness parameter $a$ yields a 2$p_{3/2}$ state that is lower in energy than the 1$f_{7/2}$ state, as was shown in Ref. \cite{HSCB10}. A similar potential with $a=0.75$ fm is used as that in Ref. \cite{HSCB10} for $^{31}$Ne. 
For the HFB calculations, a density-dependent contact pairing interaction of the surface type is utilized, in which the parameters are adjusted in order to reproduce the empirical neutron pairing gap for $^{30}$Ne \cite{YG04}.  
While the Woods-Saxon potential for the mean-field part is fixed, the pairing potential is obtained self-consistently with the contact interaction.

The top panel of Fig. \ref{ah:fig2} shows the mean square radius of the 
2$p_{3/2}$ state for $^{31}$Ne. 
In order to investigate the dependence on the single-particle energy, the depth of the Woods-Saxon well for the  $p_{3/2}$ states in $^{31}$Ne is varied. 
The dashed line is obtained with the {HF} single-particle wave function, while the solid line is obtained with the wave function for the canonical basis in the HFB calculations. 
One can see an extremely large increase of the radius of the Wood-Saxon wave function for the $p$-wave state in the limit of $\varepsilon_{\rm WS}\rightarrow 0$. In contrast, the HFB wave functions show only a small increase in radius even in the limit of $\varepsilon_{WS}\rightarrow 0$.  
This behavior persists even when contributions from other orbitals are included, as shown in the middle panel of Fig. \ref{ah:fig2}. 
Due to the pairing effect in the continuum, the HFB calculations yield a larger radius than the HF calculations for cases of $\varepsilon_{\rm WS}\le -1$ MeV.  
On the other hand, in the range of $-1$ MeV $<\varepsilon_{\rm WS}<0$ MeV the HF wave function (equivalently one quasi-particle wave function in HFB) extends significantly, while the HFB wave function 
 remains compact  due to the pairing anti-halo effect.

In the bottom panel of Fig. \ref{ah:fig2}, the average pairing gap is shown as a function of the single-particle energy $\varepsilon_{\rm WS}$.  
It is seen that the average pairing gap increases as $\varepsilon_{\rm WS}$ approaches zero. This arises because the pairing field couples with the extended wave functions of weakly bound nucleons in the self-consistent calculations. 
{That is, the pairing field extends together with the wave functions and becomes larger for a loosely bound system. }

Contrary to the above discussions of the anti-halo effect, it was claimed in Ref. ~\cite{Ham05} based on a simplified HFB model, that the pairing gap nearly vanishes in the limit of $\varepsilon\rightarrow 0$ and thus the anti-halo effect might not be realized.
In that simplified model, the pairing potential is fixed during the solution of the HFB equations, and the Fermi energy $\lambda$ is set equal to the single-particle energy $\varepsilon$ closest to the Fermi surface. 
The pairing gap of this model has been studied varying $\lambda$ so that the condition of particle number conservation is fulfilled \cite{HS12}.  
It is found that the effective pairing gap persists even in the limit of $\varepsilon\rightarrow$0, and the anti-halo effect exists clearly as in the consistent HFB model even if the pairing potential is fixed to be a constant.  
\begin{figure}[htp]
\vspace{-0.5cm}
\centering 
\includegraphics[clip,width=8.cm,bb= 0 0 400 550]{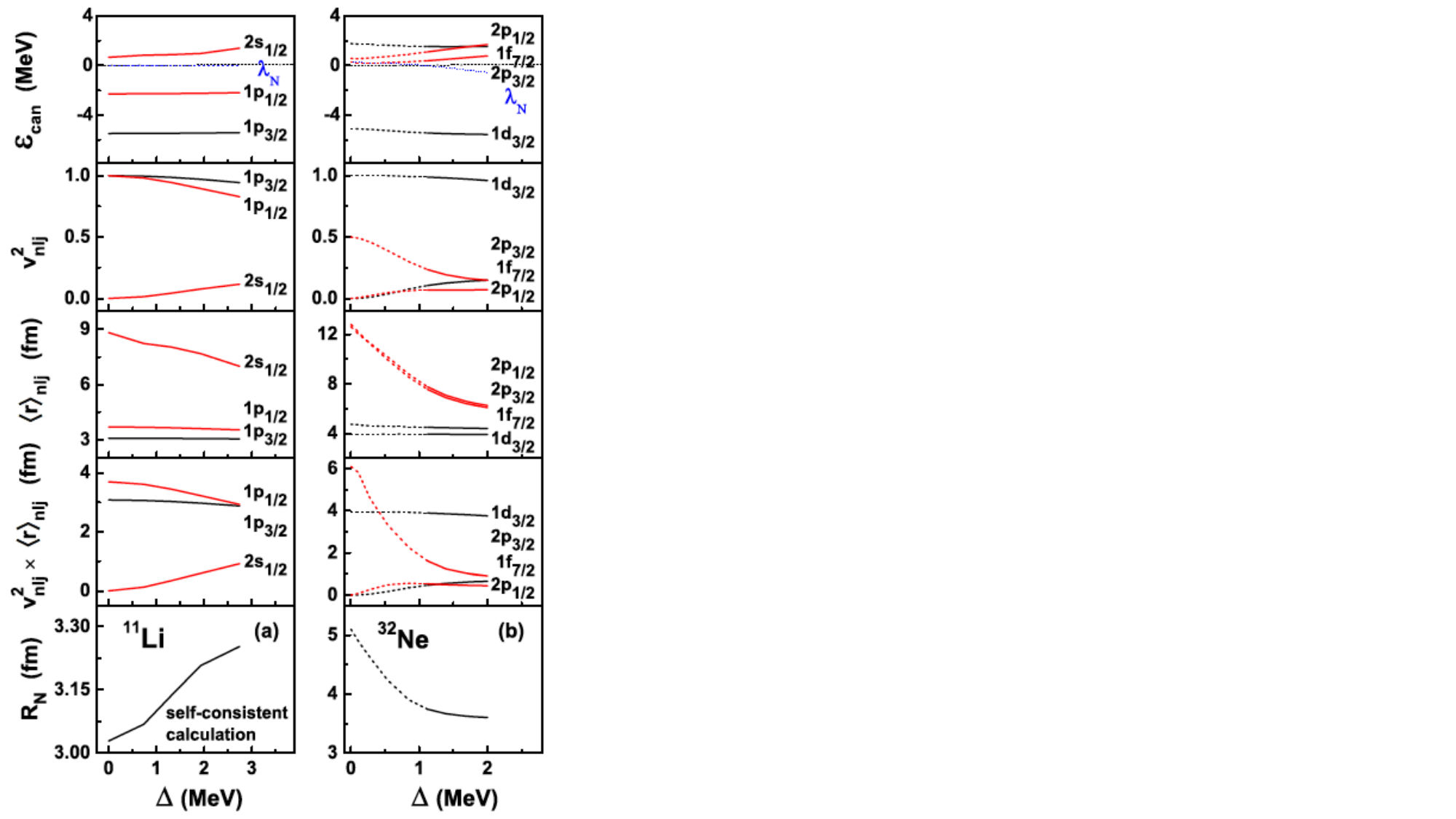}
\caption{The single-particle energy of the canonical basis, the occupation probability, the rms radius of each orbit, the rms radius multiplied by occupation probability, and the total neutron radius for $^{11}$Li and $^{32}$Ne, respectively calculated by the RCHB theory.  Reprinted figure with permission from Ref. \cite{Chen2014}. Copyright (2014) by the
American Physical Society. \label{anti-halo}}
\end{figure} 

{When the contribution from the continuum is taken into account, the so called ``anti-halo'' effect becomes more complicated and interesting. The relativistic continuum Hartree-Bogoliubov (RCHB) theory has been applied to study the effect of the continuum states on the radii of nuclei near the neutron drip line \cite{Chen2014}. } Figure \ref{anti-halo} shows the calculated single-particle energy of the canonical basis $\varepsilon_{\rm can}$, the occupation probability $v^2_{nlj}$, the rms radius of each orbit $\langle r\rangle_{nlj}$, the rms radius multiplied by occupation probability, and the total neutron radius $R_N$ for $^{11}$Li (left column) and $^{32}$Ne (right column),  respectively. 
We can see a completely opposite trend in $R_N$ for $^{11}$Li  and $^{32}$Ne as a function of the pairing gap $\Delta$ in the last rows of the figure; $R_N$ increases for larger $\Delta$ value in $^{11}$Li, while it decreases in the case of $^{32}$Ne.  In $^{11}$Li,  the occupation probability of  $2s_{1/2}$ orbit in the continuum increases due to the virtual excitation of 2$n$, which effectively increases the  $R_N$ value for larger $\Delta$.  The 2$n$ excitation to the continuum $2p_{1/2}$ state is very small, so that $R_N$ value decreases by the anti-halo effect in $^{32}$Ne.  Thus we can see a clear restoration of the anti-halo effect in $^{11}$Li, while it survives in the case of  $^{32}$Ne in contrast.

{To summarize this section, the anti-halo effect 
 was claimed firstly by a HFB calculation in Ref. \cite{BDP00}. 
The experimental evidence was found in the odd-even staggering of the reaction cross sections of Ne and Mg isotopes \cite{Takechi12,mg371}.  
Notice that the actual radius of a whole 
nucleus is determined as a consequence of the pairing anti-halo effect 
and another effect of the pairing correlation, that is, the scattering of a nucleon pair 
to higher energy continuum states ~\cite{Chen2014,Nakada18}. These two effects 
compete with each other, and 
the latter effect may become important in certain  nuclei. 
Careful examinations of pairing correlations are needed for the  study of anti-halo effect and its restoration together with further experimental information.}

\section{Soft dipole excitations in deformed halo {nuclei}}\label{sec4}
In this section,  we briefly review the study of  dipole excitation in deformed halo nuclei, focusing on the lower energy region {below} the giant dipole resonance. 

\subsection{Deformed Woods-Saxon model for dipole excitation}
 
\begin{figure}[ht]
 \centering
 \includegraphics[width=10cm]{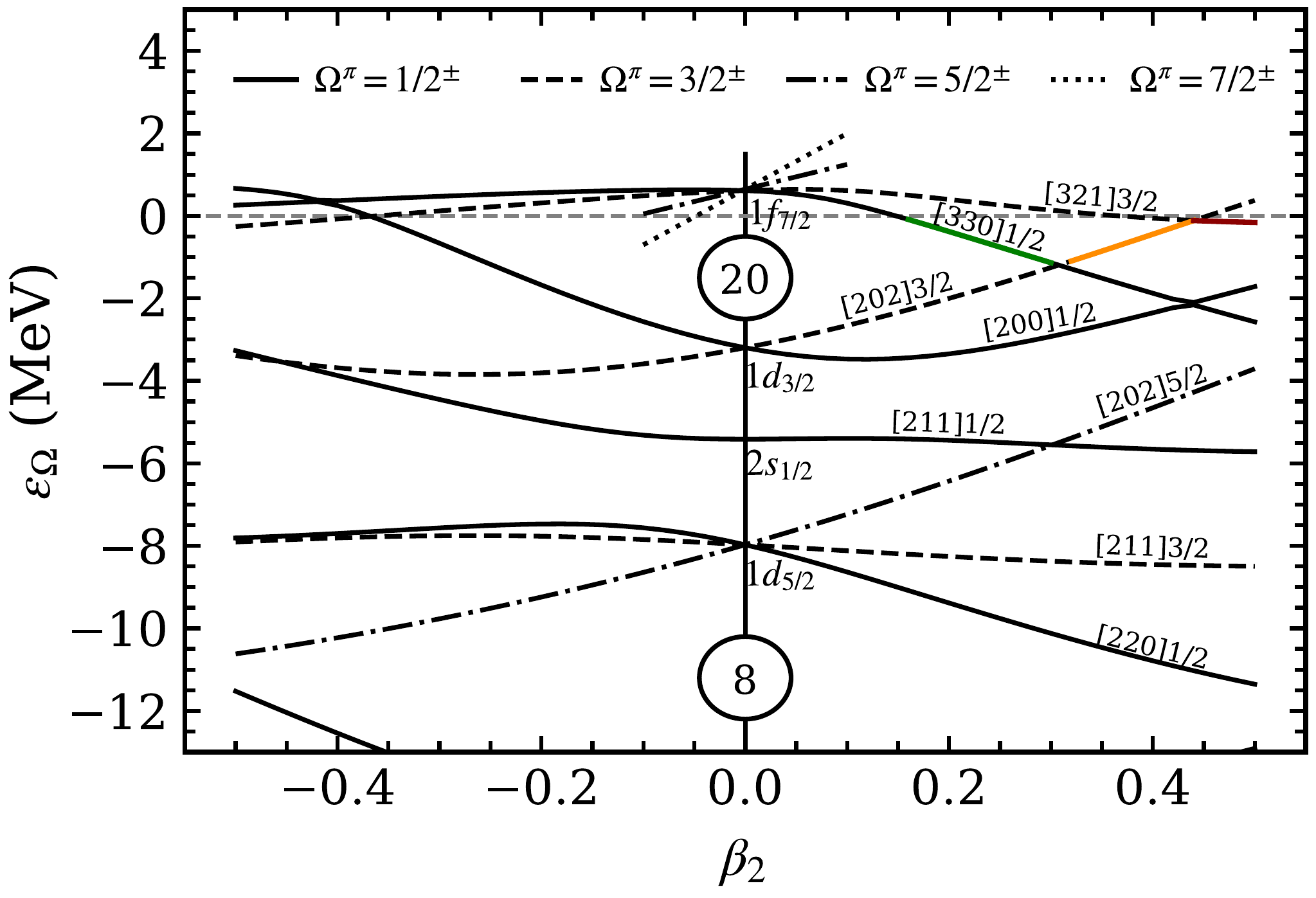}
 \caption{Single-particle levels for neutrons in deformed Woods-Saxon potentials as a function of the quadrupole deformation parameter $\beta_{2}$. The potential
depth $V_{\mathrm{WS}}$ is adjusted so that 
the binding energy of the 21st neutron of the prolate deformed nucleus $^{31}$Ne is 150 keV.
The asymptotic quantum-numbers $[Nn_z\Lambda]\Omega$ are denoted for the single-particle levels. {(Reproduced with permission from \cite{Xiao2025}. Copyright (2025) by the American Physical Society.)}
}   \label{fig-SPE}
\end{figure} 

The soft dipole excitations of the deformed halo nuclei $^{31}$Ne and $^{37}$Mg were investigated by using the deformed Woods-Saxon model in Ref. \cite{Xiao2025}, in which   detail formulas of the model are given. Figure \ref{fig-SPE} shows the calculated Nilsson diagram for the $^{30}\text{Ne}+n$ system, where the potential depth $V_{\mathrm{WS}}$ is adjusted to reproduce the one-neutron separation energy of $S_n \approx 150$ keV. The valence neutron outside the $N=20$ core occupies the negative-parity $[330]1/2$ orbit at small deformations ($0.15 < \beta_2 < 0.3$), transitions to the positive-parity $[202]3/2$ orbit in the medium deformation range ($0.3 < \beta_2 < 0.4$), and eventually enters the $[321]3/2$ orbit at large deformations ($\beta_2 > 0.4$).

To analyze the spatial characteristics of these states, the intrinsic wave functions of the Nilsson orbits are decomposed into their spherical basis components $(n \ell j)$. The $[321]3/2$ state is primarily composed of $p_{3/2}$, $f_{5/2}$, and $f_{7/2}$ waves, while the $[330]1/2$ state involves $p_{1/2}$, $p_{3/2}$, $f_{5/2}$, and $f_{7/2}$ components. In contrast, the positive-parity $[202]3/2$ orbit consists of $d_{3/2}$ and $d_{5/2}$ waves. Analysis of the wave function compositions shows that the negative-parity orbits exhibit significant contributions from the halo $p$-components, which are absent in the positive-parity $[202]3/2$ state. Furthermore, the calculated probability of the $p_{3/2}$ component in the $[321]3/2$ configuration is 32.7\%, which is in excellent agreement with the experimental value of 32\% inferred from Coulomb breakup measurements \cite{Nakamura2014}, supporting the reliability of the current model.

\begin{figure}
 \centering
 \includegraphics[width=14cm]{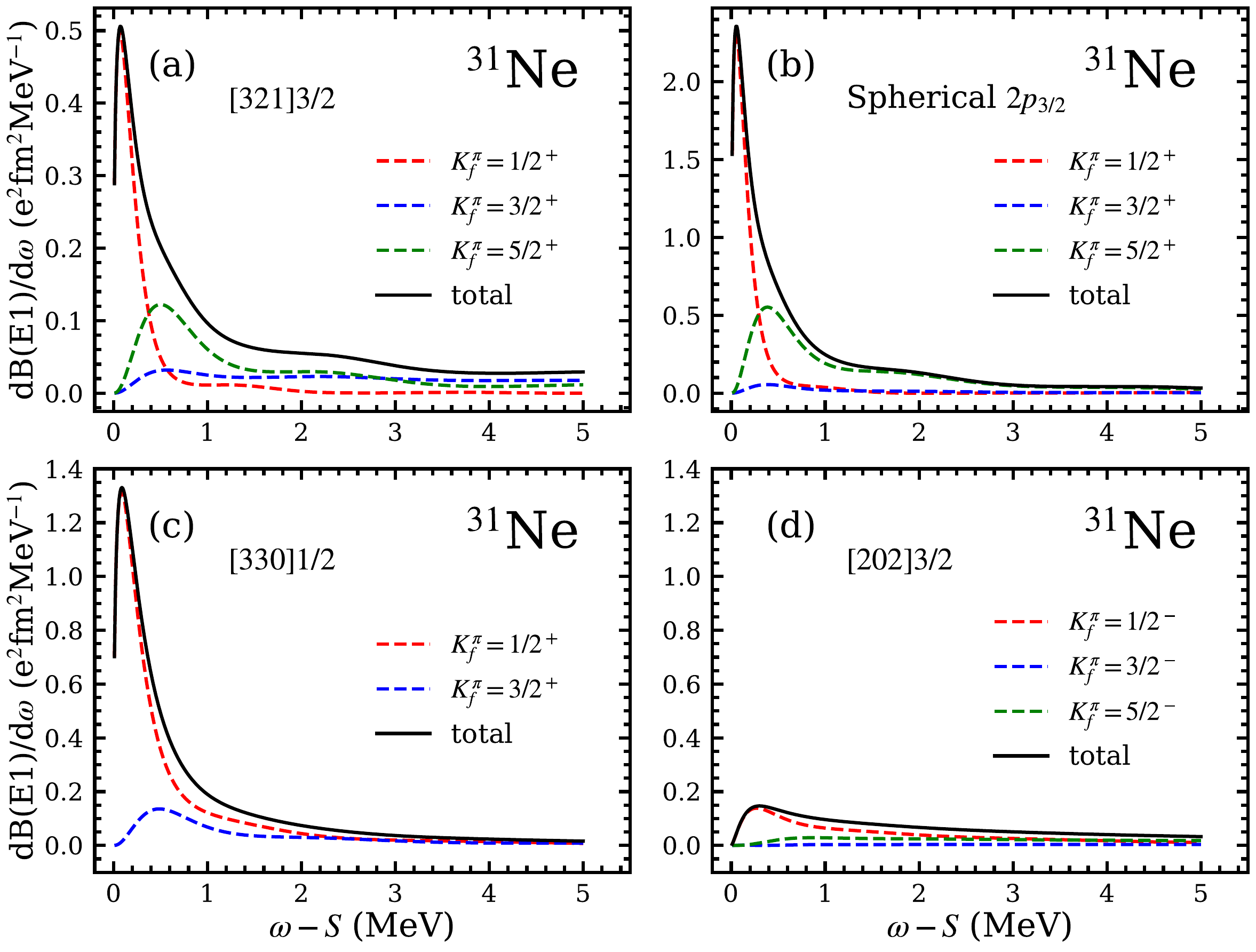}
 \caption{ (Color online) Dipole strength distributions $dB(E1)/d\omega$ calculated within the PWA as a function of excitation energy relative to the particle threshold. (a) Results for the $[321]3/2$ ($I_i^\pi=3/2^-$) ground-state configuration at $\beta_2=0.5$ with a neutron separation energy $S_n=0.15$ MeV. The $p_{3/2}$, $f_{5/2}$, and $f_{7/2}$ components are included in the initial state wave function. Final angular momenta $I_f^\pi$ are summed for each $K_f^\pi$ branch. (b) Comparison with a pure single-particle $2p_{3/2}$ halo state with $S_n=0.15$ MeV. (c) Same as (a), but for the $[330]1/2$ ($I_i^\pi=3/2^-$) configuration at $\beta_2=0.24$. (d) Same as (a), but for the positive-parity $[202]3/2$ ($I_i^\pi=3/2^+$) configuration at $\beta_2=0.32$. {Reproduced with permission from \cite{Xiao2025}. Copyright (2025) by the American Physical Society.)}}   \label{figSD}
\end{figure} 

Figure \ref{figSD} illustrates the $E1$ strength distributions ($dB(E1)/d\omega$) for $^{31}$Ne calculated within the plan wave approximation (PWA) for various initial configurations. For the $[321]3/2$ configuration at $\beta_2=0.5$, the initial state is assigned as $I_i^\pi = K_i^\pi = 3/2^-$ in accordance with experimental observations~\cite{Nakamura2014}. According to the $E1$ selection rules ($\Delta I = 0, \pm 1; \Delta K = 0, \pm 1$), the transition connects the initial state to several continuum final states: $K_f^\pi=1/2^+$ (with $I_f^\pi=1/2^+, 3/2^+, 5/2^+$), $K_f^\pi=3/2^+$ (with $I_f^\pi=3/2^+, 5/2^+$), and $K_f^\pi=5/2^+$ (with $I_f^\pi=5/2^+$). The resulting dipole response is displayed in Fig. \ref{figSD}(a), where the strengths are summed over the final angular momenta $I_f$ for each $K_f$ branch.

As shown in the figure, the $E1$ response to the $K^\pi_f=1/2^+$ state exhibits a very sharp peak just above the threshold, while the $K^\pi_f=3/2^+$ and $5/2^+$ states show relatively broad peaks at locations slightly higher than the threshold. This sharp threshold enhancement in the $1/2^+$ branch is a hallmark of soft dipole excitation, driven by the $s$-wave component in the continuum, which {feels no centrifugal barrier.} The combined contributions of these six states lead to an enhanced single-peak structure in the $E1$ strength distribution near the threshold.

Notably, compared with the response from a pure spherical $2p_{3/2}$ configuration in Fig. \ref{figSD}(b), the peak height for the deformed $[321]3/2$ configuration is quenched to only about 20\% of the spherical limit. {This reduction is a direct consequence of the deformation: in the $[321]3/2$ Nilsson orbit, the weight of the halo-favoring $p_{3/2}$ component is only 32.7\%, while the rest of the wave function consists of higher-$\ell$ components ($f$-waves) that do not contribute to the low-energy threshold enhancement.} 
 
The $E1$ response for the $[330]1/2$ configuration at $\beta_2=0.24$ is shown in Fig. \ref{figSD}(c). In this case, the final states are limited to those with $K_f^\pi=1/2^+$ and $3/2^+$. Similar to the $[321]3/2$ case, the $1/2^+$ channel dominates the threshold behavior with a sharp peak. Notably, the total $E1$ strength for $[330]1/2$ is nearly twice as large as that for $[321]3/2$. This enhancement is consistent with the larger occupation probability of the $p$-wave component (approximately 60\%) in the $[330]1/2$ orbit, further demonstrating that the soft dipole strength is a sensitive probe of the $p$-wave weight in the deformed halo ground state.

Figure \ref{figSD}(d) shows the $E1$ response from the positive-parity $[202]3/2$ orbit, which is a candidate configuration for the $N=21$ neutron in the medium deformation range ($\beta_2 \sim 0.3$). Unlike the negative-parity orbits, the dipole response here lacks a sharp threshold peak and instead displays a broad, suppressed distribution.
Due to the $K$-quantum number selection rules, there is no $s$-wave halo component involved in the transition, leading to the disappearance of the characteristic ``soft dipole'' enhancement near the threshold. 

Another deformed halo nucleus $^{37}$Mg is also studied with the same deformed Woods-Saxon model in Ref. \cite{Xiao2025}. For this system, the one-neutron separation energy is constrained to $S_n \approx 220$ keV at a large deformation of $\beta_2 = 0.46$, consistent with recent empirical data~\cite{mg37}. The valence neutron is assigned to the $[321]1/2$ Nilsson orbit, which, similar to the orbits in $^{31}$Ne, involves significant $p$-wave components. The calculated $E1$ response exhibits a prominent soft dipole nature. However, when compared with the dipole strength of a pure $p$-wave configuration, it is evident that the deformation effect partially quenches the peak intensity of the $E1$ strength distribution.

\subsection{The effect of pairing force on the distribution of dipole response}

    \begin{figure}
    \centering
    \includegraphics[width=0.5\linewidth]{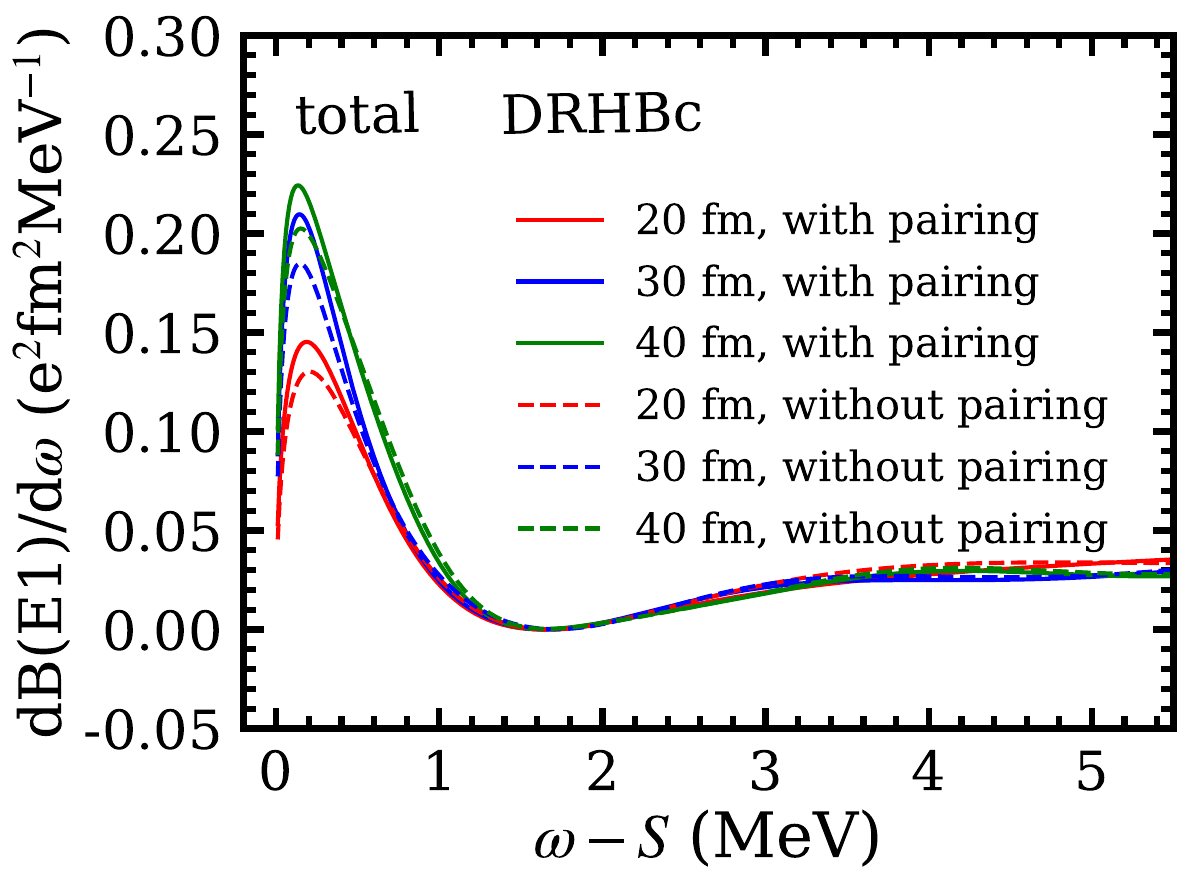}    \caption{The dipole response of $^{37}$Mg to the final state with $K^\pi_f=1/2^+$, $3/2^+$ (total) calculated using the DRHBc model as a function of the excitation energy. Different colors correspond to different radius truncations. The solid and dashed lines represent the cases with and without pairing force, respectively.}
    \label{Mg37dprDRHBc-wop}
\end{figure}

\begin{table}[ht!]
    \centering\small
\setlength{\tabcolsep}{7pt} 
    \caption{Integrated $B(E1)$ values (in $e^2\text{fm}^2$) from the threshold to $\omega-S = 1.5$, 5, and 20 MeV. Rows highlighted in yellow represent calculations including the pairing correlations, while non-highlighted rows are without the pairing correlations.}
    \label{tab:placeholder}
    \begin{tabular}{cccc}
        \hline\hline
        Box size & \multicolumn{3}{c}{Upper limit of $\omega-S$ (MeV)} \\
        \cline{2-4}
        (fm) & 1.5 & 5 & 20 \\
        \hline
        20 & 0.090 & 0.165 & 0.558 \\
        \rowcolor{yellow!30}[\tabcolsep][\tabcolsep] 20 & 0.093 & 0.160 & 0.540 \\
        \hline
        30 & 0.111 & 0.178 & 0.577 \\
        \rowcolor{yellow!30}[\tabcolsep][\tabcolsep] 30 & 0.119 & 0.182 & 0.549 \\
        \hline
        40 & 0.136 & 0.205 & 0.587 \\
        \rowcolor{yellow!30}[\tabcolsep][\tabcolsep] 40 & 0.138 & 0.204 & 0.568 \\
        \hline\hline
    \end{tabular}
\end{table}

In Sec. \ref{sec3}, we reviewed the pairing anti-halo effect on the rms radii of halo nuclei. In what other observables does the anti-halo effect manifest? In this subsection, we further examine the anti-halo effect and its restoration in the dipole response of deformed halo nuclei. Figure \ref{Mg37dprDRHBc-wop} displays the dipole response of the deformed halo nucleus $^{37}$Mg, calculated by using the DRHBc model with the PC-F1 functional, both with and without pairing correlations. {The dipole response is calculated following an approach similar to the deformed Woods-Saxon potential model~\cite{Xiao2025}, but using the ground-state bound wave functions from the DRHBc model as the initial states. In these calculations, only the large components of the initial wave functions are taken into account.} The calculation box size was varied from 20 to 40 fm in steps of 10 fm. In all cases, the dipole response with pairing correlations exhibits a higher peak intensity just above the threshold compared to the case without pairing. 

The integrated $B(E1)$ values for excitation energies $\omega-S \le$ 1.5, 5.0, and 20 MeV are summarized in Table \ref{tab:placeholder}. The results indicate a slight enhancement of the $B(E1)$ strength in the low-energy peak region ($0 < \omega-S \le 1.5$ MeV) due to pairing. Conversely, a small quenching effect is observed in the higher energy region ($5 < \omega-S \le 20$ MeV). These findings suggest that in the dipole response of the deformed halo nucleus $^{37}$Mg, the pairing anti-halo effect is largely canceled by the strong coupling to the continuum, a phenomenon similar to that observed for the neutron rms radius of $^{11}$Li.


To summarize this section, the study of deformed halo was claimed in a theoretical paper in 1997 \cite{MISU1997}, and has been one of the most challenging subject experimentally~\cite{TANIHATA2013215,Nakamura2014,mg371,plbzhang2024} and theoretically \cite{MISU1997,PhysRevC.54.1617,PhysRevC.69.041306,PEI200629,NAKADA200847,pei132,Meng_2015,Nakada18,2023-CNPC56-IPP}.
So far, two solid experimental evidence was found in $^{31}$Ne and $^{37}$Mg by the observation of soft dipole excitations in RIKEN RIBF experiments.  
The soft dipole excitations of these two deformed halo nuclei have been investigated by using the deformed Woods-Saxon model. It has been pointed out that the $E1$ strength near the neutron threshold is dependent on the configuration of the valence neutron. In this work, we have also explored the anti-halo effect of pairing on the dipole excitations and revealed the important contribution of continuum coupling which leads to more pronounced dipole strength near the threshold. 

\section{Summary}\label{sec5}
We have discussed two related topics of nuclear halo phenomena in this Contribution to the Halo40 Proceedings. The first one is the pairing anti-halo effect on the neutron radius and 
its restoration due to the coupling to the continuum, i.e., the virtual neutron excitations to the continuum configurations.  The second one is the soft dipole excitation of deformed halo nuclei.   We pointed out the importance of HFB and HB in continuum theories to take into account properly the halo nature of extended wave function in the calculations of neutron radius and also the soft  dipole excitation of halo nuclei.  It was shown that the anti-halo effect is very sensitive to the continuum coupling induced by the Bogoliubov type quasi-particle, which cancels largely the anti-halo effect on the neutron radius.  
   
   The soft dipole excitations of deformed halo nuclei $^{31}$Ne and $^{37}$Mg were discussed in the deformed Woods-Saxon model, whose potential depth is adjusted to reproduce the empirical separation energies of these nuclei.  We found  that the 
   shape peak just above the threshold is created by the halo effect and its strength can be used to identify the size of deformation and the Nilsson configuration of deformed halo orbit.    Furthermore, the self-consistent DRHBc model was applied to study the effect of 
   pairing correlation on the soft dipole excitation in $^{37}$Mg.  It was shown  that the anti-halo effect on the dipole response is largely canceled by the coupling to the continuum in the DRHBc model.

\begin{acknowledgments} 
This work is supported by the National Natural Science Foundation of China (Grant Nos. 12447101, 12347139, 12375118, 12435008, and W2412043), the National Key R\&D Program of China (Grant Nos. 2023YFA1606500 and 2024YFE0109800) and the CAS Strategic Priority Research Program (Grant Nos. XDB34010100 and XDB0920000). The results described in this paper are obtained on the High-performance Computing Cluster of ITP-CAS and the ScGrid of the Supercomputing Center, Computer Network Information Center of Chinese Academy of Sciences. 
\end{acknowledgments}

 
\bibliography{ref.bib}
\end{document}